\documentclass[%
 reprint,
superscriptaddress,
 amsmath,amssymb,
 aps,
]{revtex4-2}

\usepackage{graphicx}
\usepackage{dcolumn}
\usepackage{bm}
\usepackage{hyperref}
\usepackage{natbib}
\usepackage{cleveref}
\usepackage{mathbbol}
\usepackage{xcolor}
\usepackage[normalem]{ulem}

\begin{document}


\title{Lattice real-time simulations with learned optimal kernels}

\author{Daniel Alvestad}
\author{Alexander Rothkopf}
\affiliation{%
Department of Mathematics and Physics,University of Stavanger, 4021 Stavanger, Norway
}%

\author{D\'enes Sexty}
\affiliation{%
Institute of Physics, NAWI Graz, University of Graz, Universit\"atsplatz 5, Graz, Austria
}%


\begin{abstract}
We present a simulation strategy for the real-time dynamics of quantum fields, inspired by reinforcement learning. It builds on the complex Langevin approach, which it amends with system specific prior information, a necessary prerequisite to overcome this exceptionally severe sign problem. The optimization process underlying our machine learning approach is made possible by deploying inherently stable solvers of the complex Langevin stochastic process and a novel optimality criterion derived from insight into so-called boundary terms. This conceptual and technical progress allows us to both significantly extend the range of real-time simulations in 1+1d scalar field theory beyond the state-of-the-art and to avoid discretization artifacts that plagued previous real-time field theory simulations. Limitations of and promising future directions are discussed.
\end{abstract}

\maketitle


\section{Introduction}

What unites many of the pressing open questions in modern physics, irrespective of whether they relate to eV (condensed matter-), MeV (nuclear-) or GeV (particle physics) energy scales, is the need to access the dynamics of strongly correlated quantum many-body systems in Minkowski time. Concretely, as e.g.~outlined in the recent Snomass community review \cite{davoudi_report_2022,nachman_quantum_2021}, an \textit{ab-initio} understanding of transport properties of nuclear matter at high temperature and density, as well as the scattering of showers of high energy partons still remain out of reach of state-of-the-art analytic and numerical Monte-Carlo methods. First principles insight into real-time transport of non-relativistic fermions \cite{hangleiter_easing_2020} and their interaction with gauge fields is a key puzzle piece in understanding high-temperature superconductivity (
e.g. in the Hubbard model \cite{qin_hubbard_2022}). Fission and fusion dynamics \cite{bender_future_2020}, too, 
remain currently out of reach of fully ab-initio field-theoretic approaches, requiring model input.

Vital ab-initio insight into the static (thermodynamic) properties of strongly correlated many-body systems has been achieved in the past through Monte-Carlo simulations of Feynman's path integral \cite{smit_introduction_2002}. These numerical techniques rely on analytic continuation to an unphysical Euclidean time. In turn the extraction of 
relevant real-time dynamics becomes an ill-posed inverse problem \cite{rothkopf_bayesian_2022}, which severely affects the accurate 
determination of central quantities of interest: from transport coefficients  \cite{Brambilla:2022xbd,Altenkort:2023oms}, to in-medium decay rates \cite{Kim:2018yhk,Larsen:2019zqv}, to vacuum parton distribution functions \cite{Karpie:2019eiq,Liang:2019frk}. Developing a \textit{direct simulation approach} in Minkowski time is thus called for.

Direct simulations of real-time dynamics suffer from the so-called \textit{sign problem} \cite{gattringer_approaches_2016,pan_sign_2022}. Feynman's path integral is formulated as a sum over field configurations weighted by a complex phase. A minute signal emerges from the sum of a vast number of almost cancelling phases, overwhelming otherwise efficient Markov-chain sampling based approaches. Some sign problems have been proven \cite{troyer_computational_2005} to belong to the class of \textit{NP-hard computational problems}, which entails that no generic solution method in polynomial time exist on a classical computer. Various approaches have been proposed to tackle the sign problem, such as reweighting (RW), extrapolation (EX) \cite{de_forcrand_constraining_2010,braun_imaginary_2013,braun_zero-temperature_2015,guenther_qcd_2017}, density of states (DS) \cite{wang_efficient_2001,langfeld_density_2012,gattringer_density_2015}, tensor networks (TN)  \cite{orus_tensor_2019,meurice_tensor_2022}, Lefschetz thimbles (LT) \cite{rom_shifted-contour_1997,cristoforetti_new_2012,Alexandru:2020wrj} and complex Langevin (CL) \cite{Klauder:1983nn,Parisi:1984cs} . They all propose a \textit{system agnostic} recipe to the estimation of observables in the presence of a sign problem. Without a system specific component, each of these methods are \textit{destined to eventually fail}, be it that their computational cost scales unfavorably when applied to systems in realistic volumes in 3+1 dimensions (RW,TN,DS,LT) or they suffer from convergence to an incorrect solution (CL).

Quantum computing offers a different angle of attack to the sign problem \cite{preskill_quantum_2018}, as in principle it can compute the unitary time evolution of a spin-system. The mapping of a realistic field theory to spin systems remains an open challenge, especially if gauge degrees of freedom are involved \cite{nachman_quantum_2021}. The necessity to derive a Hamiltonian for implementation on a quantum computer, to date, requires truncation of the continuous state space and in the case of photons and gluons fixing to a particular gauge. It is acknowledged in the quantum computing community (see e.g.~\cite{bauer_quantum_2022,Dalzell:2023ywa}) that with near-future noisy intermediate scale devices, many physical systems of interest remain too complex to be modelled with quantum circuits. 

Therefore, progress in the short-term requires innovation among \textit{system specific} real-time simulation techniques on classical computers.
\vspace{-0.25cm}
\section{Real-time Complex Langevin}
\vspace{-0.25cm}
Here we build upon the complex Langevin approach, which is one of the complexification strategies to the sign-problem, with similarities but important differences to contour deformations (LT) (see discussion in \cite{aarts_remarks_2014,Alvestad:2022abf}). In conventional stochastic quantization \cite{damgaard_stochastic_1987,namiki_stochastic_1992} one proves that the expectation values of a Euclidean quantum field theory $\langle{\cal O}\rangle(\tau)=\int {\cal D}\phi_E {\cal O}(\phi_E) {\rm exp}[-S_{\rm E}(\phi_E)]$ can be reproduced by simulating a stochastic process in an additional 
Langevin time $\tau_L$ direction, using the Langevin equation $\partial_{\tau_L}\phi(\tau_L,\tau)=-\delta S_{\rm E}/\delta\phi_E +\eta(\tau_L,\tau)$ with Gaussian noise 
$\langle \eta(\tau_L,\tau) \eta(\tau_L^\prime,\tau^\prime)\rangle=2\delta(\tau_L-\tau_L^\prime)\delta(\tau-\tau^\prime)$.

\begin{figure}
\centering
\includegraphics[scale=0.7, trim= 0cm 0.2cm 0 0.5cm, clip=true]{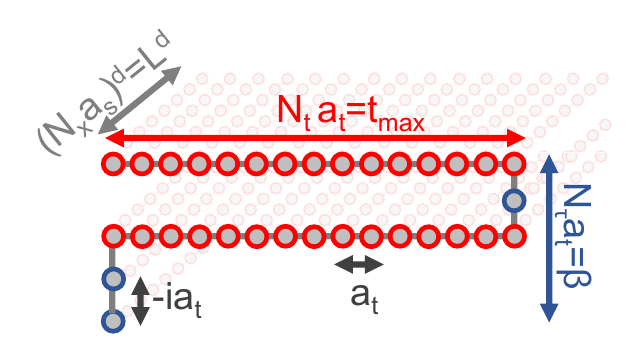}
\caption{Geometry of discretized (d+1) dimensional scalar field theory on the Schwinger-Keldysh contour. Here we use $N_t=32$, $N_\tau=4$ and $N_x=8$ and $a_tm=1/10, a_sm=2/10$.}\label{fig:SKTthermdiscr}
\vspace{-0.3cm}
\end{figure}

Quantum field theory with a mixed initial density matrix in Minkowski time constitutes an initial value problem and is formulated on the Schwinger-Keldysh contour ${\cal C}_{\rm SK}$ with a forward- and backward branch, housing the fields $\phi_1$, $\phi_2$ respectively
\begin{align}
    &\langle {\cal O}\rangle (t,{\bf x})=\hspace{-0.1cm}\int\hspace{-0.1cm}d[\phi_1^{(i)},\phi_2^{(i)}] \rho(\phi_1^{(i)},\phi_2^{(i)})\hspace{-0.1cm}\int_{\phi_1^{(i)}}^{\phi_2^{(i)}}\hspace{-0.2cm}{\cal D}\phi\,{\cal O}e^{iS_{{\cal C}_{\rm SK}}}\\
    &\overset{\rho_\beta}{=}\hspace{-0.1cm}\int\hspace{-0.1cm}{\cal D}\phi_E e^{-S_E} \int_{\phi_E^{(f)}}^{\phi_E^{(i)}}\hspace{-0.2cm}{\cal D}\phi\,{\cal O}e^{iS_{{\cal C}_{\rm SK}}}=\int_{{\cal S}_{\rm SKt}}\hspace{-0.2cm}{\cal D}\phi\,{\cal O}e^{iS_{{\cal C}_{\rm SKt}}}\label{eq:PIonSK}
\end{align}
In a thermal system at $T=1/\beta$, we have $\rho_\beta\sim{\rm exp}[-\beta H]$ and sampling over initial conditions can be written as a path integral on a compact imaginary time domain of length $\beta$, connecting the real-time branches as closed contour ${\cal C}_{\rm SKt}$. We parametrise ${\cal C}_{\rm SKt}$ in the complex time plane with the real contour parameter $\gamma$: $t(\gamma)$. 
Cauchy's theorem allows us to deform the integration contour and we choose the convention sketched in \cref{fig:SKTthermdiscr}, where the downward portion of ${\cal C}_{\rm SKt}$ is divided into two pieces at $t=t_{\rm max}$ and $t=t_0$.

Naive CL proposes to complexifiy the field d.o.f.~$\phi_c=\phi_R+i\phi_I$ and to carry out the following coupled stochastic process \cite{namiki_stochastic_1992} for $\phi_R$ and $\phi_I$ in Langevin time $\tau_L$
\begin{align}
     \frac{d\phi_c}{d\tau_L} = i\frac{\delta S[\phi_c]}{\delta \phi_c} + \eta(\tau_L,x) \label{eq:CLeom},
\end{align}
estimating observables $\langle {\cal O}\rangle$ from the ensemble average over analytically continued observables $\langle {\cal O}\rangle[\phi_c]$. While significant progress has been made in application of CL in various model systems \cite{Seiler:2017wvd,berger_complex_2021} and even to the theory of strong interactions at finite Baryon-chemical potential \cite{Sexty:2013ica,Sexty:2019vqx,Scherzer:2020kiu,Attanasio:2022mjd}, the simulation of real-time dynamics so far has been hampered by various hurdles: 
divergencies (runaways) and convergence to incorrect solutions as the real-time
extent of the contour is increased \cite{Berges:2005yt,Berges:2006xc,Berges:2007nr}.

The runaway problem leads to a break down of the numerical solver for the Langevin equation when the process explores regions of the complexified manifold far from the origin.
It has been shown in ref.~\cite{Alvestad:2021hsi}, that it can also be understood
as a consequence of the stiffness of the non-linear CL dynamics.
This practical problem is solved by either using an adaptive 
stepsize control \cite{Aarts:2009dg}, or through the use of inherently stable implicit discretization schemes, such as 
Euler-Maruyama, for \cref{eq:CLeom} \cite{Alvestad:2021hsi}. The inherent regularization provided by the implicit scheme also makes it possible to directly simulate with CL on the real-time axis of the ${\cal C}_{\rm SKt}$ contour, without tilt.

Important insight into the convergence properties of CL have been gained in \cite{Aarts:2011ax,Nagata:2016vkn} through 
analysis of the relation between the real probability distribution $P[\phi_R,\phi_I]$ sampled by \cref{eq:CLeom} and the complex Feynman weights ${\rm exp}[iS]$ in \cref{eq:PIonSK}. Connection is made via the real Fokker-Planck operator $L$ and its complex generalization $\cal L$. For CL to correctly reproduce expectation values, the sampled distribution $P[\phi_R,\phi_I]$ must fall off sufficiently fast in $\phi_I$, to enable integration by parts. At the same time the spectrum of $\cal L$ must have negative real parts. Based on this insight, criteria for correct convergence have been developed: 
boundary terms \cite{boundaryterms1,boundaryterms2}, 
and two improvements to complex Langevin have been proposed: gauge cooling \cite{Seiler:2012wz}, where gauge freedom is exploited to keep the d.o.f. close to the original real-valued manifold and dynamic stabilization \cite{Aarts:2016qhx}, which introduces an additional drift term into \cref{eq:CLeom}. The drawback of the latter is that the new drift term is non-holomorphic and thus at odds with the proof of convergence of CL and it might introduce a bias in the results.

It is long known \cite{namiki_stochastic_1992} that real Langevin can be modified by a kernel $K$, without changing its stationary distribution. This freedom has been exploited to improve autocorrelation in Euclidean theories \cite{Batrouni:1985jn}. In complex Langevin one may introduce a complex kernel $K=K[\phi_c;\tau_{\rm L}]$ \cite{Soderberg:1987pd,Okamoto:1988ru,Okano:1991tz}. $K$ must be a holomorphic function and be factorizable as $K=H^T H$, but can otherwise be an arbitrary (matrix) function of the fields. It can encode 
transformations \cite{Aarts:2012ft} such as 
in coordinates, contour deformations or redefinition of variables. The general kernelled CL evolution equation for discretized spatial coordinates $\phi_c(x_j)=\phi_c^j$ reads 
\begin{equation}
    \hspace{-0.1cm}\frac{d\phi_c^j}{d\tau} = \left[ i K_{jk}(\phi_c)\frac{\partial S(\phi_c)}{\partial \phi_c^k} + \frac{\partial K_{jk}(\phi_c)}{\partial \phi_c^k} \right] + H_{jk}(\phi_c) \eta_k.\label{eq:CLeomK}
\end{equation}
In the past, a few system specific transformations have been found that soften (see e.g. \cite{levy2021mitigating, Boguslavski:2022dee}) or even avoided (see e.g. \cite{Okamoto:1988ru,Okano:1991tz,Aarts:2012ft}) the sign problem in model systems (see also reformulation strategies e.g. \cite{Chandrasekharan:1999cm,DelgadoMercado:2011sqg,Kloiber:2013rba}). However for realistic systems, success has been limited and no systematic recipe is known to extend results from simpler systems. This study instead uses machine learning (ML) techniques to \textit{systematically learn optimal kernels}, based on \textit{system specific prior information}.

\section{Machine-learning assisted kernelled Langevin}

Our machine learning strategy for kernelled Langevin is inspired by reinforcement learning (RL) \cite{sutton2018reinforcement}. RL underlies recent advances in diverse fields: beating computer games or steering autonomous vehicles. It is based on an agent, endowed with a set of limited actions, placed in a predefined environment. Success of the agent is encoded in a cost/policy functional defined from environment variables and the internal state of the agent. A mathematical representation of the actions of the agent allows the use of differential programming techniques \cite{baydin2018automatic} to evaluate the gradients of the cost functional w.r.t. those actions. Challenges are the robust detection of failure modes of the agent, and the trade-off between generality of the actions of the agent and learning efficiency.

Specifying to real-time simulations, we define our agent as the controller of the kernel $K$, which allows it to explore the abstract space of stationary distributions of the stochastic process \cref{eq:CLeomK}. A crucial ingredient is our use of \textit{system-specific prior information} to define the cost functional  $\mathbb{p}[K]$, used to assess the success of CL convergence. As was shown e.g.~in \cite{Alvestad:2022abf} the failure of convergence of CL on the SK contour occurs globally, i.e.~it affects correlators on all branches. In a thermal setting, time translation invariance requires equal-time correlation functions to be constant on the whole complex time-contour. Conventional simulations in the Euclidean domain in addition give access to their values, as well as to the Euclidean unequal-time correlators. Deviations from this prior knowledge are easily assessed within the CL simulations. To test for successful convergence, we deploy $\mathbb{p}[K]=\Big\{\sum_{ij}( {\rm Re}\langle \phi_c^2(\gamma_i)\rangle-\langle \phi^2\rangle_{\rm HMC})(C^{\rm Re})^{-1}_{ij}( {\rm Re}\langle \phi_c^2(\gamma_j)\rangle-\langle \phi^2\rangle_{\rm HMC})+( {\rm Im}\langle \phi_c^2(\gamma_i)\rangle)(C^{\rm Im})^{-1}_{ij}({\rm Im}\langle \phi_c^2(\gamma_j)\rangle)\Big\}$, which consists of two likelihood terms involving the covariance matrices $C^{\rm Re/Im}$ of the CL equal-time correlators ${\rm Re/Im}\langle \phi_c^2\rangle$. It thus assesses the constancy and agreement with apriori known values. $\mathbb{p}[K]$ makes reference to expectation values involving the fields $\phi_c$ and thus implicitly depends on $K$.
To obtain robust gradients w.r.t. the entries of $K$ we must take derivatives over the whole stochastic dynamics. While adjoint \cite{cao2003adjoint} and shadowing methods \cite{wang2014least} are promising to compute 
gradients, we find from 
the CL Lyapunov exponents \cite{alvestadPhD} that the 
dynamics actually becomes chaotic, degrading the performance of conventional differential programming techniques.


Instead we use a low-cost optimization functional $\mathbb{l}[K]$, which provides gradients $\nabla_K\mathbb{l}[K]$ that significantly reduce the values of the actual cost functional $\mathbb{p}[K]$. We find that\vspace{-0.2cm}
\begin{align}
    \mathbb{l}[K]=\int d^dx d\gamma {\rm Im}[\phi_c(\tau_L,\gamma,{\bf x})]^2,\label{eq:lowcost}
\end{align}
proposed in \cite{Lampl:2023xpb}, offers the best performance in minimizing $\mathbb{p}[K]$, compared to earlier choices in \cite{Alvestad:2022abf}. This improvement relies on its relation to boundary terms and to the recently derived improved correctness criterion in \cite{Seiler:2023kes}.

We restrict ourselves to the simplest type of a field- and $\tau_L$ independent kernel. Note that even though the optimization functional may contain non-holomorphic terms, the kernel does not and thus \cref{eq:CLeomK} is compatible with the proof for correct convergence. The potentially costly Jacobian $\delta K/\delta \phi_c$ also does not need to be computed. 

\vspace{-0.25cm}
\section{Numerical results}
\vspace{-0.25cm}

Let us apply our machine learning assisted complex Langevin approach to thermal scalar field theory in $1+1$ dimensions with $m=1$ and a quartic self-coupling $(\lambda/4!)\phi^4$ with $\lambda=1$ at $\beta m=4/10$, a benchmark also used in \cite{Alexandru:2017lqr}. The field is discretized on the SK contour sketched in \cref{fig:SKTthermdiscr} with $N_t=32$ points along the real-time branches each and $N_\tau=4$ steps along the imaginary time direction. The spatial dimension is resolved with $N_x=8$ points. As lattice spacing we use $a_sm=2/10$ and a finer $a_tm=1/10$, to avoid discretization artifacts from the corners of the SK contour. The discretized action is identical with the one of Ref.~\cite{Alexandru:2017lqr}.  (code available at \cite{CodeDA})

\begin{figure}
\centering
\includegraphics[scale=0.4,trim= 1cm 0.7cm 0.5cm 0cm, clip=true]{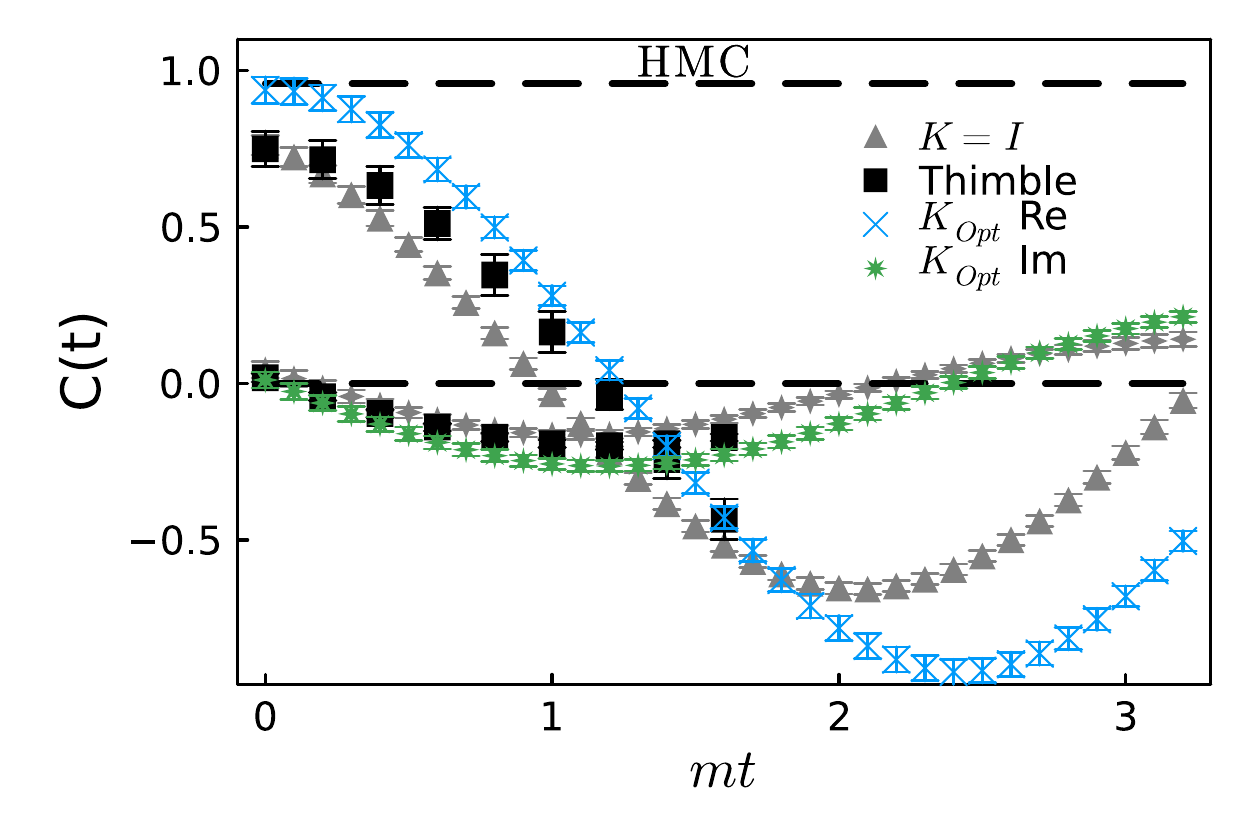}
\caption{${\rm Re[C]}$ and ${\rm Im[C]}$ in 1+1d field theory from naive (gray triangle, stars) and optimal kernel CL (blue crosses, green stars) with $N_t=32$. Result from contour deformation \cite{Alexandru:2017lqr} as black squares with $N_t=8$. The value of the correlator at $t=0$ from hybrid Monte-Carlo is given as gray solid line.} \label{fig:unequalRT} \vspace{-0.2cm}
\end{figure}

Using adaptive step size with maximum Langevin step $d\tau_L=10^{-3}$, we simulate at a real-time extent of $t_{\rm max}=3.2$, which lies deep in the region where naive CL $(K=I)$ fails to converge correctly, as shown by the gray triangles in \cref{fig:unequalRT}, denoting the real- and imaginary part of the unequal time, momentum zero, correlator $C(t)=\langle \phi_c(t,p=0)\phi_c(0,p=0)\rangle$. The failure manifests in the deviation of $C(0)$ at $K=I$ from the value of the equal-time correlation function $F(\gamma)=\langle \phi_c(\gamma,p=0)\phi_c(\gamma,p=0)\rangle$ at $\gamma=0$. Its values are known from conventional HMC simulations and indicated by the black dashed line (in 1+1d, $F(0)$ only carries a minute lattice spacing dependence).

We parameterize a fully dense, constant, complex kernel via $H=A+iB$ with real matrices $A$ and $B$, each of which have $[(2N_t+N_\tau)N_x]^2$ entries. Learning of the optimal kernel starts from the trivial choice $H = \mathbb{1}$, i.e. $K=H^T H = \mathbb{1}$. The resulting stiff dynamics is solved with an implicit Euler-Maruyama integrator for which we use the implementation in the \texttt{DifferentialEquations.jl} library \cite{rackauckas2017differentialequations} of the \texttt{Julia} language. After a Langevin time of $\tau_L^{\rm opt}= 5$ we compute the gradient of the discrete \cref{eq:lowcost} using the autodiff capability of \texttt{Julia}. Based on the \texttt{Adams} algorithm with learning rate $r_l=10^{-3}$ the entries of $A$ and $B$ are iteratively updated reducing the initial 
$\mathbb{p}[K=\mathbb{1}]\approx 1200$ to $\mathbb{p}[K_{\rm opt}]\approx7.2$. 
Observables for $K_{\rm opt}$ are obtained from 
three streams with $\tau_L^{\rm obs}=5000$.

The central result of our study, the unequal-time correlation function $C(t)$ from optimal learned kernels in 1+1d is shown as blue crosses ${\rm Re}[C]$ and green stars ${\rm Im}[C]$ in \cref{fig:unequalRT}. We reach a real-time extent of at $t_{\rm max}=3.2$ which is twice that was previously achieved in the literature using contour deformations \cite{Alexandru:2017lqr}. Note that our simulation results for ${\rm Re}[C](0)$ and ${\rm Im}[C](0)$ both agree with $F(0)$ from HMC, given by given by the dashed black lines. 

We emphasize that the advantageous scaling properties, that our approach inherits from CL, enables us to deploy a finer grid here. Implicit methods do not pose a problem, as highly optimized implementations of solvers are readily available. Since we restrict ourselves to field independent kernels so far, we avoid the need for Jacobians, whose computational cost is a central limiting factor for contour deformation methods. 

\begin{figure}
\centering
\includegraphics[scale=0.35,trim= 1cm 0.8cm 0.5cm 0cm, clip=true]{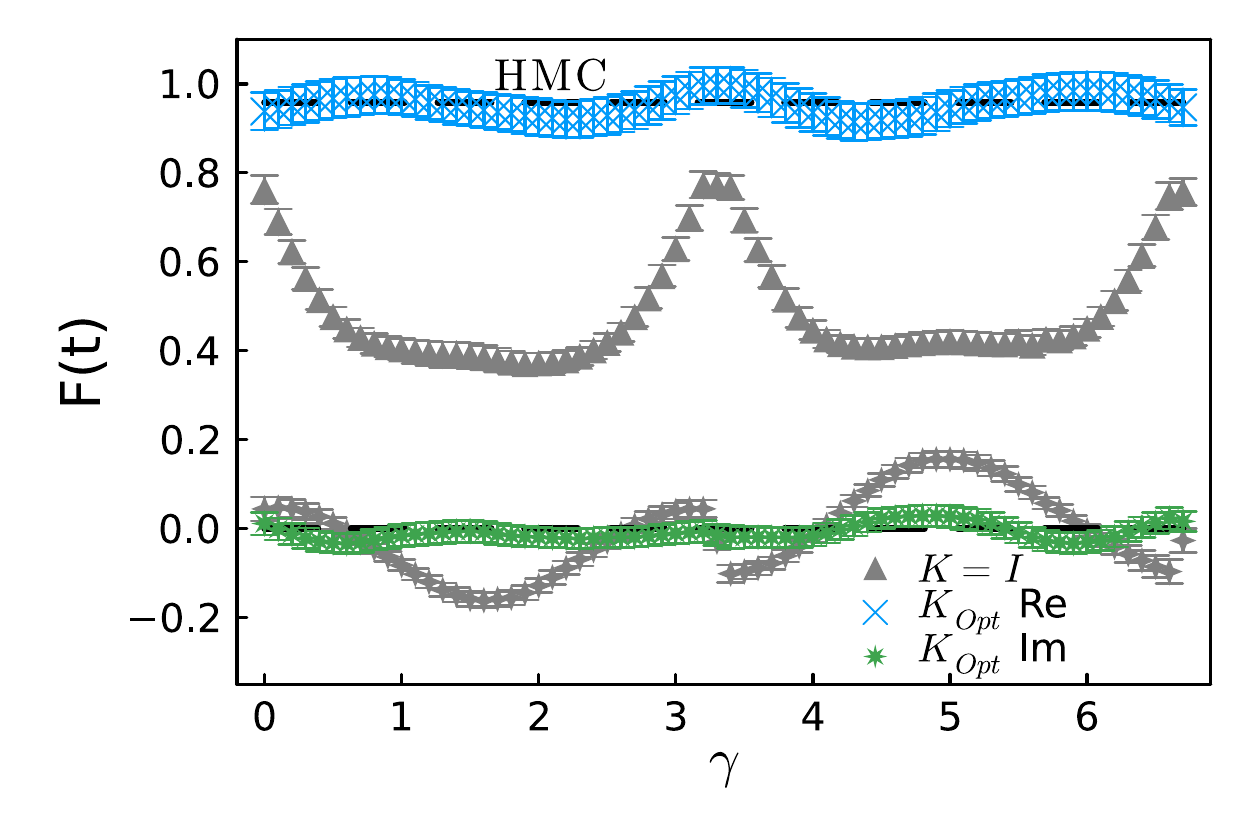}
\caption{${\rm Re}[F](\gamma)$ and ${\rm Im}[F](\gamma)$ in 1+1d scalar field theory for naive CL (gray triangle, rhombes) and for optimal learned kernels (blue crosses, green stars).} \label{fig:equalSK}\vspace{-0.2cm}
\end{figure}

We showed in \cite{Alvestad:2022abf} that the equal-time correlator $F(\gamma)$ is more difficult to reproduce in CL than the unequal-time correlator $C(t)$ and thus plot the former against the contour parameter $\gamma$ in \cref{fig:equalSK}. Note that both ${\rm Re}[F](\gamma)$ and ${\rm Im}[F](\gamma)$, while showing minute oscillations, agree with the apriori known values (gray dashed lines). 
The optimal kernel CL results are in stark contrast to naive CL $(K=I)$, for which $F(\gamma)$ (gray triangles, rhombes) clearly deviates from the HMC. This crosscheck provides convincing support for the correctness of convergence. 

\begin{figure}
\centering
\includegraphics[scale=0.35, trim= 0cm 0cm 0cm 0cm, clip=true]{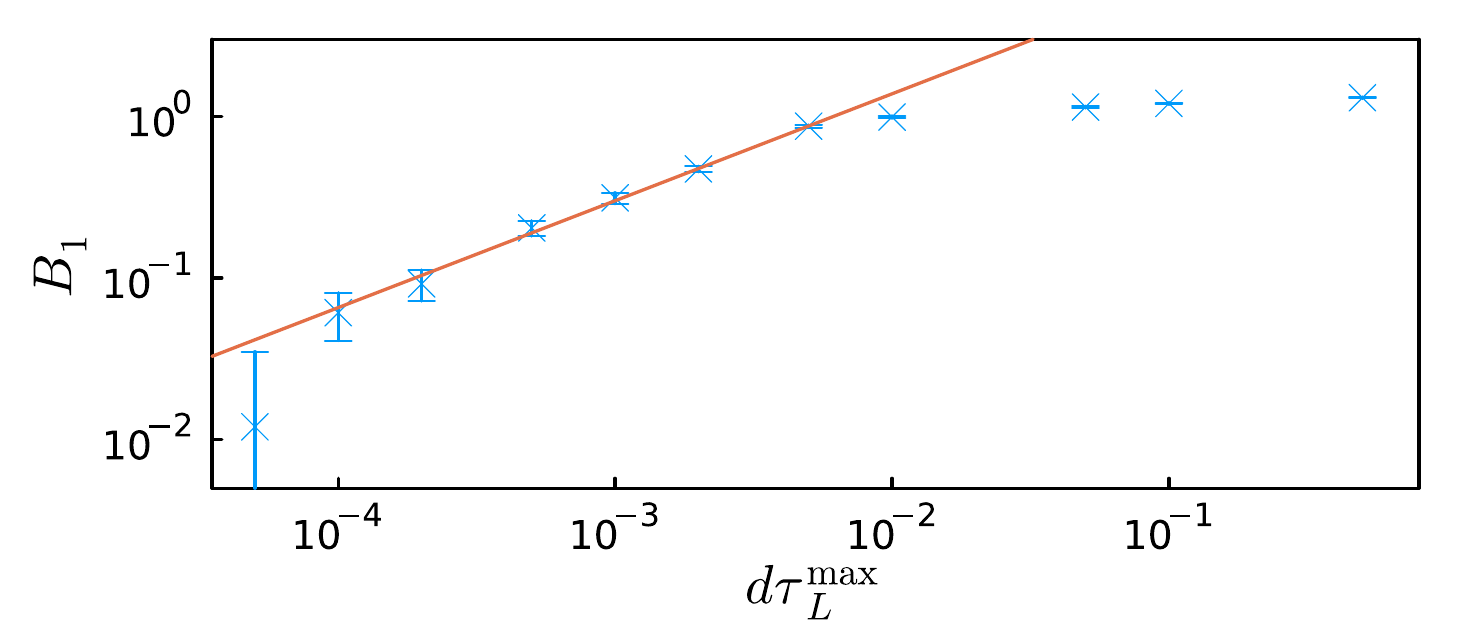} 
\vspace{-0.4cm}
\caption{Magnitude of the dominant boundary term $B_1$, based on the $\langle \phi^2\rangle$ observable. Note the best fit powerlaw dependence on $(d\tau_L^{\rm max})^{0.66}$ consistent with correct convergence.} \label{fig:B1}\vspace{-0.4cm}
\end{figure}

Further support for correct convergence is provided from the absence of  boundary terms, such as $B_1$ (see \cite{boundaryterms1,boundaryterms2}) for the $\langle \phi^2\rangle$ observable. As shown in \cref{fig:B1} $B_1$ exhibits a powerlaw dependence on $d\tau_L$, consistent with vanishing boundary terms in the continuum limit.

\begin{figure}\centering
    \includegraphics[scale=0.098, trim= 2.8cm 0.8cm 1.5cm 1.5cm, clip=true]{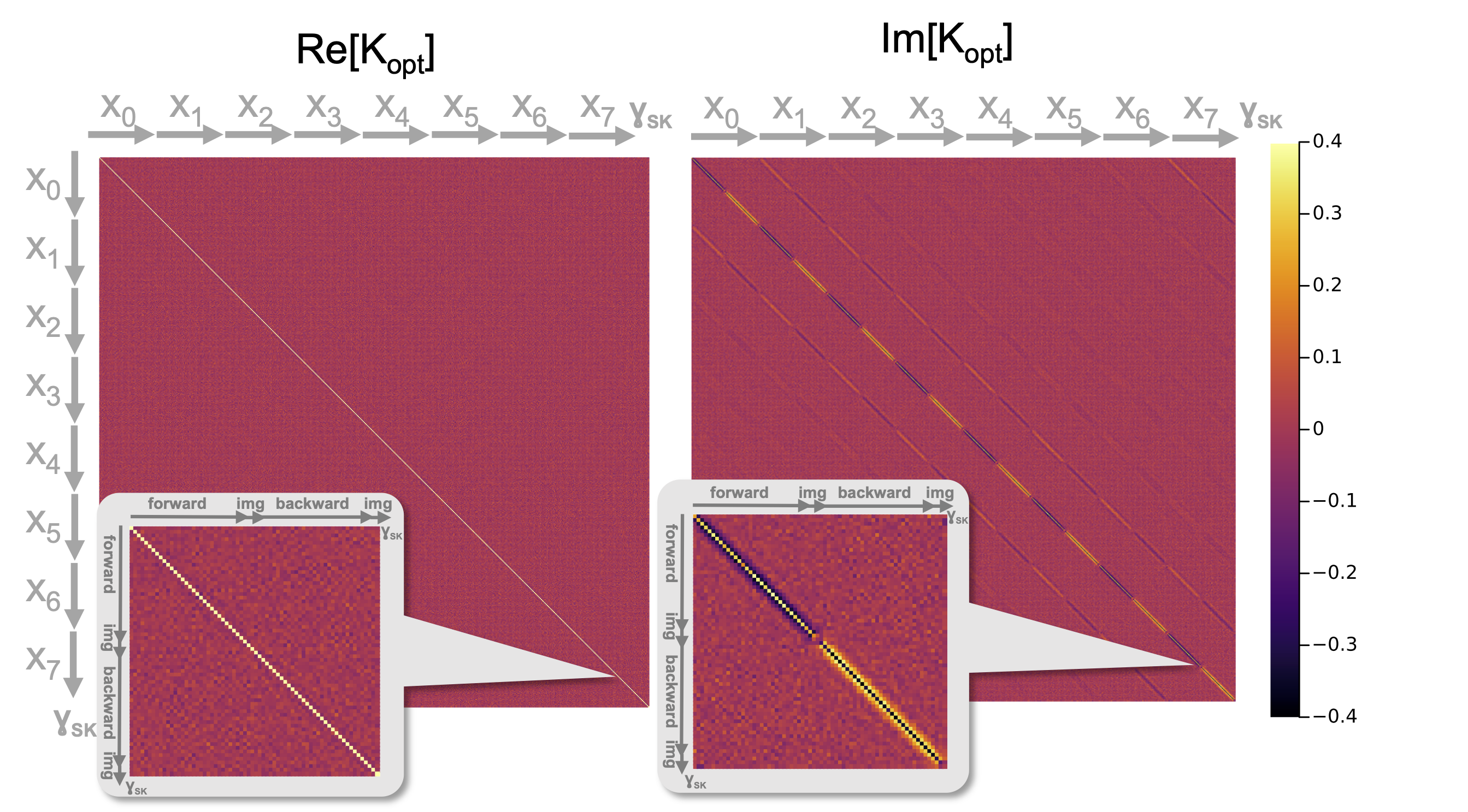}
    \caption{(top) Entries of the real- (left) and imaginary part (right) of the optimal kernel underlying \cref{fig:unequalRT,fig:equalSK}. (bottom insets) zoom-in of (left) ${\rm Re}[K_{\rm opt}]$ and (right) ${\rm Im}[K_{\rm opt}]$ on a single spatial slice with the subsections of ${\cal C}_{\rm SKt}$ labelled on gray arrows. (see text for detailed discussion)}
    \label{fig:1p1dOptimalKernel}\vspace{-0.2cm}
\end{figure}

In the top row of \cref{fig:1p1dOptimalKernel} we plot the values of the optimal learned kernel underlying \cref{fig:unequalRT,fig:equalSK}. Color coding resolves values $|K_{\rm opt}|\leq0.4$, sufficient for all but the 
diagonal entries of ${\rm Re}[K_{\rm opt}^{\rm diag}]\approx1.025$. Each pixel corresponds to a component of $K_{\rm opt}$ that connects two space-time points, ordered such, that in each spatial slice $x_i$ denoted by a gray arrow, the parameter $\gamma$ traverses the full contour ${\cal C}_{\rm SKt}$. ${\rm Re}[K_{\rm opt}]$ is dominated by the diagonal alone, while ${\rm Im}[K_{\rm opt}]$ shows a distinct banded structure in space with diminishing amplitude farther away from the diagonal, which represents the spatially non-local nature of the transformation implemented by $K_{\rm opt}$. In the two insets we highlight the behavior of the kernel in a single spatial slice, where the corresponding sections of the SK contour connected by the kernel entries are indicated by the gray arrows. Similar to the results in 0+1d, we find a characteristic finite difference-like behavior, where entries of opposite sign on the sub and supradiagonal accompany those on the diagonal, indicating a Fourier filter.

Our ML approach is able to identify a much simpler structure than what a naive extension of 
the free theory kernel suggests (c.f. \cite{Alvestad:2022abf}), a testament to the efficacy of the 
learning strategy, which holds 
potential for analytic insight into convergence restoring transformations.

While success of the ML approach is encouraging, it too will fail at larger real-time extents. Establishing an exact range of validity is work in progress. One reason is that we only optimize based on the low-cost functional $\mathbb{l}[K]$ and not directly on $\mathbb{p}[K]$. The development of robust gradient estimators for chaotic stochastic systems (see e.g. NILSAS for the Lorenz system \cite{alvestadPhD}) is called for. Another reason is the thimble structure of the theory (see also \cite{Mou:2019tck}). One-d.o.f. models tell us that parameters exist, in which a field-dependent kernel is needed to capture the physics of multiple contributing thimbles. In that case the kernel can be systematically expanded via a rational approximation involving $\phi_c$, limiting computational cost. Transfer learning between kernels of different expressivity will be key to keep cost in check.

In summary we have presented a machine learning approach to direct real-time simulations on the lattice, in which \textit{system-specific} prior information is incorporated into CL via iterative ML of an optimal field independent kernel. Using a novel low-cost functional for the computation of gradients for learning, we achieve efficient convergence in 1+1d field theory to at least twice the real-time extent previously accessible. Due to the efficiency of the approach, we can access fine grids to avoid discretization artifacts affecting previous studies. Work is ongoing to extend the results to realistic $(3+1)d$ and we explore the inclusion of field dependent kernels. 
\begin{acknowledgments}
D.~A.~ and A.~R. thank Rasmus Larsen for helpful discussions and gladly acknowledge support by the Research Council of Norway under the FRIPRO Young Research Talent grant 286883. 
D.~S.~ acknowledges the support of the Austrian Science Fund (FWF) through the Stand alone Project P36875.
The study has benefited from computing resources provided by  
UNINETT Sigma2 - the National Infrastructure for High Performance Computing and Data Storage in Norway under project NN9578K-QCDrtX "Real-time dynamics of nuclear matter under extreme conditions". Some parts of the numerical calculations 
where performed on GSC, the computing cluster of the University of Graz.

\end{acknowledgments}

\bibliographystyle{apsrev4-1}
\bibliography{RealTimeCLML}

\end{document}